\documentclass[fleqn,10pt,keys]{wlscirep}
\usepackage[utf8]{inputenc}
\usepackage[T1]{fontenc}

\newcommand{\size}{1\textwidth}

\title{Anomalous Fano Resonance in Double Quantum Dot System Coupled to Superconductor}

\author[1,*,+]{Jan Bara\'nski}
\author[1,{\#}]{Tomasz Zienkiewicz}
\author[1,{\$}]{Magdalena Bara\'nska}
\author[2,*,{\&}]{Konrad Jerzy Kapcia}
\affil[1]{Polish Air Force University, ul. Dywizjonu 303, PL-08521 D{\c e}blin, Poland}
\affil[2]{Institute of Nuclear Physics, Polish Academy of Sciences, ul. W. E. Radzikowskiego 152, PL-31342 Krak\'{o}w, Poland}

\affil[*]{Correspondence and requests for materials should be addressed to J.B. (email: {j.baranski@law.mil.pl}) or K.J.K. (email: konrad.kapcia@ifj.edu.pl)}

\affil[{{+}}]{email: j.baranski@law.mil.pl; ORCID ID: \url{http://orcid.org/0000-0002-0963-497X}}
\affil[{{\#}}]{email: t.zienkiewicz@law.mil.pl;  ORCID ID: \url{http://orcid.org/0000-0002-7111-2590}}
\affil[{{\$}}]{email: m.baranska@law.mil.pl;  ORCID ID: \url{http://orcid.org/0000-0003-1964-6373}}
\affil[{{\&}}]{email: konrad.kapcia@ifj.edu.pl; ORCID ID: \url{https://orcid.org/0000-0001-8842-1886}}


\begin{abstract}
We analyze the influence of  {a} local pairing on  {the} quantum interference in nanoscopic systems.
As a model system we choose the double quantum dot coupled to one metallic and one superconducting electrode in  {the} T-shape geometry. 
The analysis is particularly valuable for systems containing coupled objects with considerably different broadening of energy levels. 
In such systems,  {the} scattering of itinerant electrons on a discrete (or narrow) energy level gives rise to the Fano-type interference. 
Systems with induced superconducting order, along well understood Fano resonances, exhibit also another features on the opposite side of the Fermi level. 
The lineshape of these resonances differs significantly from their reflection on the opposite side of the Fermi level, and their origin was not fully understood. 
Here, considering  {the} spin-polarized tunneling model, we explain  {a} microscopic mechanism of  {a} formation of these resonances and discuss the nature of their uncommon lineshapes.
We show that  {the} anomalous Fano profiles originate solely from  {the} pairing of nonscattered electrons with scattered ones. 
We investigate also the interplay of each type of resonances with the Kondo physics and discuss the resonant features in differential conductivity. 
\end{abstract}

\begin{document}

\flushbottom
\maketitle
%
%
\thispagestyle{empty}

\section*{Introduction}

\begin{figure}
\centering
\includegraphics[width=\size]{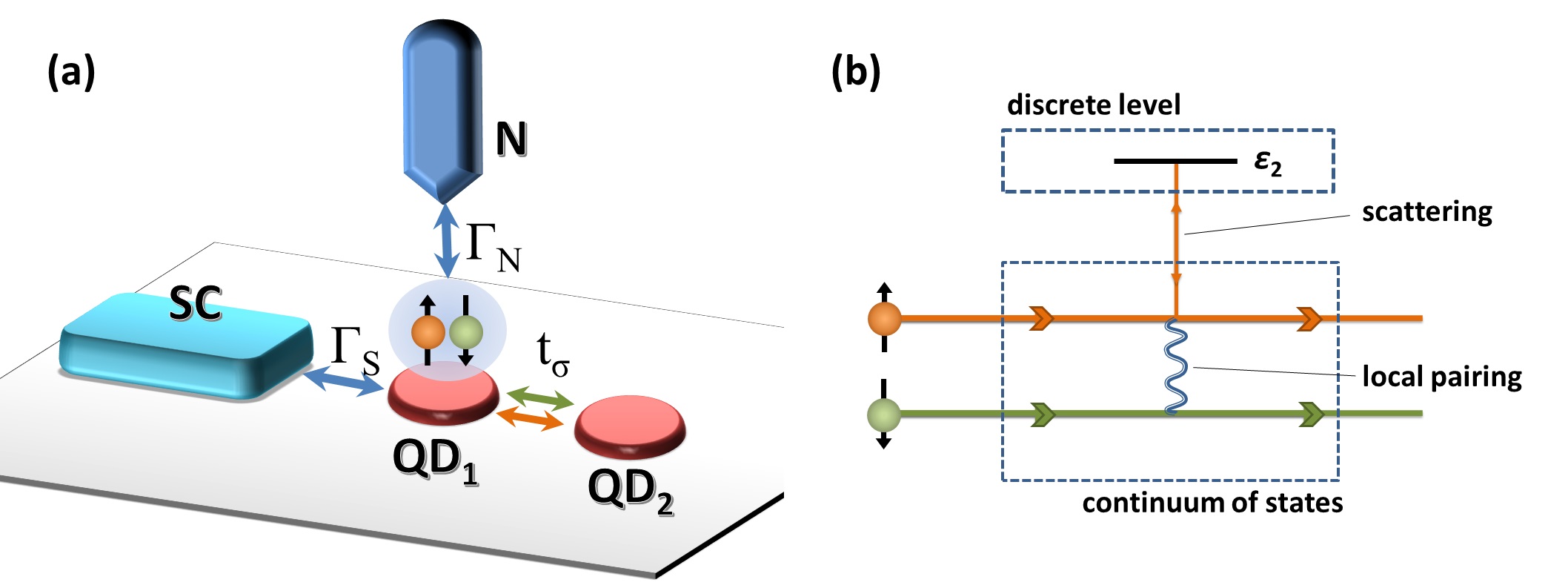}
\caption{ {(a)} The schematic illustration of the analyzed system. 
It consists of two quantum dots (QD$_1$ and QD$_2$).
QD$_1$ (interfacial one) is directly coupled with superconducting (SC) and normal metal (N) electrodes with spin-independent couplings $\Gamma_{S}$ and $\Gamma_N$, respectively. 
Coupling $t_\sigma$ between QD$_1$ and QD$_2$ (side one) is spin-dependent  {[cf., Eq.~(\ref{eq:1.hamiltonian})]}.
{(b) The schematic illustration of the scattering processes occurring in the strongly spin-polarized  tunneling model ($t_{\uparrow}\neq 0 $ and $t_{\downarrow} = 0$). 
Orange (green) arrows indicate the propagation paths for $\sigma=\uparrow$ ($\sigma=\downarrow$, respectively) electrons.
Only spin-$\uparrow$ electrons can directly scatter on side dot QD$_2$ (vertical orange arrows), whereas spin-$\downarrow$ electrons are paired with them (a blue spring represents the local pairing $\Gamma_S$) and scattered indirectly.} 
} 
\label{fig:1.system}
\end{figure}

Impurities or nanoobjects like quantum dots (QDs) hybridized to superconductors (SC) adopt some SC properties via proximity effects.
As a consequence, the ground state of a QD is represented by either single particle state $\left| \uparrow \right\rangle$, $ \left| \downarrow \right\rangle$ or a superposition of empty and doubly occupied states $u_{d} \left| 0 \right\rangle + v_{d}\left| \uparrow \downarrow \right\rangle$ \cite{Lee_NatNano_2013,Bauer_2007,BaranskiJPCM2013}. 
The fingerprints of this local pairing can be observed in the Andreev spectroscopy as two quasiparticle peaks \cite{DeaconPRL2010,DeaconPRB2010,Lee_NatNano_2013}. 
Currently, dynamic development in fabrication of complex nanodevices on the top of SC substrate allows to construct SC-based systems built of multiple QD's \cite{ZitkoPRL2018}, quantum rings \cite{BURLAKOV2017,Gurtovoi2016}, monoatomic chains \cite{Kisiel2016}, gate-controlled carbon nanotubes (CNT) \cite{Kong_Science_1999,NovotnyPRL2006}, multiwall CNT quantum dots \cite{Schonenberger2004CNTQD}, modified Aharonov-Bohm rings with a QD embedded within one of the ring's arms \cite{YacobyPRL_1995}, SQUID interferometers with a gate-controlled CNT quantum dots \cite{Cleuziou2006} or quantum dots connected to Rashba chains \cite{Deng2016}.
In such systems, the various paths for electron propagation give rise to quantum interference effects. 
Therefore, deep understanding of mutual relations between  {the} proximity induced pairing and  {the} quantum interference is highly demanded.  
A classic model to analyze such relations consists of a QD (QD$_1$) coupled directly to (i) one metallic electrode and (ii) one superconducting electrode as well as  side-coupled to second QD (QD$_2$) [a schema of the system is shown in Fig.~\ref{fig:1.system}(a)]. 
In such system, the main charge transport between electrodes leads directly through the central quantum dot (i.e., QD$_1$), Fig.~\ref{fig:1.system}(b). 
Additional path includes  {the} electron hopping between the central dot (QD$_1$) and the side dot (QD$_2$). 
Different paths for electron transport overlap giving rise to quantum interference effects. 
As the interfacial quantum dot (i.e., QD$_1$) is connected to superconducting reservoir, scattering on a side level is accompanied by  {the} local pairing.
In metal-hybrid structures interference patterns can be observed in  {the} spectral function and transport characteristics as asymmetric Fano features emerging at energies equal to  {the} energy level of the side dot(s) \cite{FERNANDES201898,TorioPRB2002,Orellana2003,Takazawa_DQDT,KangDQD,BulkaPRL2001,GUEVARA2006,WeymannPRB2014}.
In the presence of local pairing, two resonant structures emerge simultaneously on both sides of the Fermi level \cite{BaranskiPRB2011,BaranskiPRB2012,Glodzik2017}.
{A} shape of the feature located at the energy of the side dot resembles the ordinary Fano resonance.
However, a structure on the opposite side of  {the} Fermi level seems to diverge from the ordinary Fano profile \cite{BaranskiPRB2011,BaranskiPRB2012}.
Although a particle-hole mixing of states rationalizes an appearance of two resonances instead of just one, astonishing difference in their profiles is intriguing.
One could even argue whether shape of additional resonance should be referred to as  {the} Fano-like.

The Fano-like profiles have been reported in numerous works in various fields of physics including atomic \cite{PhysRevLett.10.516,FANO1968,PhysRevA.55.1544}, molecular \cite{Linn1983} physics, photonics \cite{Limonov2017,Fan2002}, plasmonics \cite{Giannini2011,Lukyanchuk2010}, electron-phonon interaction\cite{PhysRevB.8.4734,PhysRevLett.68.2834,PhysRevB.74.212301,PhysRevLett.74.470,BaranskiChinPhysB2015,Vinod2018}, microwave physics \cite{FRinOptics,Micro}, metamaterials and nonlinear optics \cite{Smson2010,Lukyanchuk2010,Jung2019}, ultra cold gases \cite{Li2019}, or nuclear physics \cite{Bhatia1984,Plffy2007,Orrigo2006,Bledsoe1971}. 
The Fano resonances turn out to be particularly relevant also for nanoscale physics.
In various systems, in which nanoobjects with different broadenings of energy levels are tunnel-coupled, similar resonances appear on the background of the Breit-Wigner resonance (or, equivalently, the Lorentz distribution). 
Such Fano-like resonances were predicted and observed, e.g., in double \cite{Tanaka2005,TrochaPRB2007,TrochaPRB2014,WeymannPRB2014,Balatsky2007} and triple \cite{He_2013,Glodzik2017} quantum dots systems in various configurations \cite{Tanaka2005}.
 {The} asymmetric resonances were also predicted in  {``\emph{bridge}''} realization, where two electrodes were tunnel coupled to a single QD and, additionally, to each other directly \cite{BulkaPRL2001,HofstetterPRL2001}. 
In such a realization, the Fano effect arises as a result of interference of waves traveling directly between electrodes with a localized state. 
It was predicted that the Fano resonances appearance in a similar configuration can enhance the effectiveness of a Cooper spliter device \cite{Gong_2016}.

Recently, it has been noticed that the Fano resonances can be useful in indicating the existence of the Majorana bound states \cite{Wang_2016_QDM,Orellana2018,Schuray}. 
In systems, in which the quantum dot is weakly connected to the Rashba chain,  {the} scattering on the Majorana zero mode (MZM) suppresses  {the} local density of states (LDOS) of QD only by one half \cite{Schuray,BaranskiJPCM2017}. 
This is because electrons scattered by the Majorana quasiparticle change their phase only by the fraction of $\pi$, while in the ordinary Fano  {effect}, we observe $0-\pi$ phase shift features. 
Taking this into account, one can distinguish scattering of electrons on the MZM from scattering on the topologically trivial zero energy states. 
Rich interplay of  {the} Fano resonances with strong correlations effects was analyzed by number of authors both in metallic \cite{BulkaPRL2001,Frank2015, Madhavan, Stefanski} and superconducting environment \cite{TanakaPRB2008,TanakaPRB2010, SasakiPRL2009, ZitkoPRB2010, BaranskiPRB2011, Baraski2012APA, Baraski2012APA2}. 
Among others, it was found that suppression of the Kondo state by its coexistence with the Fano antiresonance reveals a novel Fano-Kondo resonance \cite{SasakiPRL2009,ZitkoPRB2010}. 
A pedagogical review of the Fano resonances in nanoscale physics was done by A. E. Miroshmishenko in Ref. \cite{MiroshnichenkoRMP2010}.
Interplay of the Fano resonance itself with local pairing was less widely explored.
P. Orellana and coworkers \cite{OrellanaPhysLettA2013,Orellana_2017} analyzed configuration with one quantum dot placed between two metallic electrodes and side dot coupled only to  {the} SC electrode. 
In such realization, scattering on narrow quasiparticle states gives rise to two Fano-like features on background of single particle broad level.

In this work, we present the analysis of the local pairing for electrons scattered on  {the} side structure. 
In the considered model, the SC electrode is connected directly to  {the} interfacial dot, thus quasiparticle states are considered as broad continuum while scattering occur on  {the} dot decoupled from the SC environment  {(cf. Fig. \ref{fig:1.system})}.
We discuss origin of appearing resonances and reveal the microscopic mechanism of their formation.
We analyze the shape of obtained resonances by comparing them with the Fano profiles and calculate characteristic Fano factors such as asymmetry parameter. 
We also discuss the interplay of each resonant feature with  {the} Kondo resonance and inspect the appearance of resonant features in differential conductivity.

\section*{Formulation of the problem}

A heterojunction depicted in  {Fig.~\ref{fig:1.system}(a)} can be modeled by the Anderson impurity Hamiltonian in the following form 
\begin{equation}
\label{eq:1.hamiltonian}
\hat{H}=\hat{H}_{N} + \hat{H}_{S} + \sum_{\beta=N,S} \hat{H}_{T \beta}+ \sum_{i=1,2} \hat{H}_{QDi} + \hat{H}_{t},
\end{equation}
where 
$\hat{H}_{N} = \sum_{k, \sigma} \xi_{k N} \hat{c}^{\dagger}_{k \sigma N} \hat{c}_{k\sigma N}$ 
represents the metallic reservoir and 
$\hat{H}_{S}=\sum_{k, \sigma} \xi_{kS} \hat{c}^{\dagger}_{k \sigma S} \hat{c}_{k \sigma S} - \sum_k ( \Delta \hat{c}^{\dagger} _{k\uparrow S} \hat{c}^{\dagger}_{-k \downarrow S}+h.c.)$  
refers to $s$-wave superconducting electrode. 
Electron energies $\xi_{k\beta}$ are measured 
with respect to chemical potentials $\mu_{\beta}$ ($\beta=N,S$). 
Two quantum dots connected with spin-dependent  interdot hoppings  
$t_{\sigma}$ ($\sigma = \uparrow , \downarrow$) 
are represented by the following terms:
\begin{equation}
\hat{H}_{QDi}= 
\sum_{i, \sigma} \epsilon_{i} \hat{d}^{\dagger}_{i \sigma} \hat{d}_{i \sigma} + 
\sum_{i} U_{i} \hat{n}_{i \uparrow} \hat{n}_{i {\downarrow}}, 
\qquad \textrm{and} \qquad
\hat{H}_{t} = 
\sum_{\sigma} t_{\sigma} \left( \hat{d}^{\dagger}_{1 \sigma} \hat{d}_{2 \sigma} + 
\hat{d}^{\dagger}_{2 \sigma} \hat{d}_{1 \sigma}\right),
\end{equation}
where $\epsilon_{i}$ is the energy level of $i$-th quantum dot, $U_{i}$ stands for intra-dot Coulomb interactions ($i=1,2$), and $\bar{\sigma}$ denotes the spin opposite to $\sigma$ (e.g., $\downarrow=\bar{\uparrow}$). 
The hybridization of  {the} interfacial (i.e., $i=1$) quantum dot to the external reservoirs ($\beta=N,S$) is given by
\begin{equation}
\hat{H}_{T \beta}= 
\sum_{k, \sigma} V_{k \beta \sigma} \left(\hat{c}_{k \beta \sigma} \hat{d}^{\dagger}_{1 \sigma} + h.c.\right) .
\end{equation}
It is useful to introduce the wide band limit constant coupling strength
between the interfacial dot and both reservoirs: $\Gamma_{\beta}= \pi \sum |V_{k \beta}|^2 \delta(\omega-\xi_{k})$.
In the deep superconducting atomic limit ($\Delta \gg \Gamma_{S}$), the influence of superconducting electrode on interfacial quantum dot QD$_1$ is reduced to  {the} induced local pairing. 
A problem of  {the ``proximized''} quantum dot was widely explored by many authors \cite{Bauer_2007, YamadaPRB2011,  Tanaka2007PSJ, MengPRB2009, Donabid2008, Donabidowicz2008A} 
including ourselfs \cite{BaranskiJPCM2013, Domaski2016, Zapalska2014, Domaski2014}.
In such conditions, the Hamiltonian of  {the} interfacial dot coupled to  {the} SC reservoir (i.e., $H_{QD_{1}}+H_{S}+H_{TS}$) can be expressed by 
\begin{equation}
\hat{H}_{prox} = \hat{H}_{QD_{1}} + \hat{H}_{S} + \hat{H}_{TS} =   
 \sum_{\sigma} \epsilon_{1} \hat{d}^{\dag}_{1 \sigma} \hat{d}_{1 \sigma} - 
{\Gamma_{S}}\left( \hat{d}_{1 \uparrow} \hat{d}_{1 \downarrow} + h.c.\right) + 
  U_{1} \hat{n}_{1 \uparrow} \hat{n}_{1 \downarrow}.
\end{equation}   
To gain a clear picture of the interplay between interference effects and the local pairing we will mostly focus on noncorrelated regime, i.e., $U_{1}=U_{2}=0$ (excluding the section, where correlations are studied explicitly). 
Information on spectral properties and Andreev transmittance is encoded in particular Green's functions 
$G_{j}(t_1,t_0)=-i \theta(t_1-t_0) \langle \lbrace \hat{\Psi}_{j \sigma}(t_1), \hat{\Psi}_{j \sigma}^{\dagger}(t_0) \rbrace \rangle$ of $4 x 4$ matrix  
$\hat{\Psi}_{j \sigma}^{\dagger} \equiv (\hat{d}^{\dagger}_{j \sigma}, \hat{d}_{j \sigma})$, 
$\hat{\Psi}_{j \sigma} \equiv (\hat{\Psi}_{j \sigma}^{\dagger})^{\dagger}$. 
In the present work, we assume spin-dependent interdot hopping $t_{\sigma}$, therefore, Green's functions for each spin component are not identical and for each index $\sigma= \uparrow , \downarrow$ these functions need to be calculated separately. 
In the equilibrium conditions the equation of motion technique \cite{haug2008} yields the following expression for a Fourier transform of the retarded Green function matrix for  {the} interfacial quantum dot (QD$_1$):
\begin{equation}
\check{G}_{1 \sigma}(\omega)= 
\left( \begin{array}{cc}  
\langle\langle \hat{d}_{1 \sigma} \hat{d}^{\dagger}_{1 \sigma} \rangle \rangle &
\langle\langle \hat{d}_{1 \sigma} \hat{d}_{1 \bar{\sigma}} \rangle \rangle \\
\langle\langle \hat{d}^{\dagger}_{1 \bar{\sigma}} \hat{d}^{\dagger}_{1 \sigma} \rangle \rangle & 
\langle\langle \hat{d}^{\dagger}_{1 \bar{\sigma}} \hat{d}_{1 \bar{\sigma}} \rangle \rangle \\ 
\end{array}\right) =
\left( \begin{array}{cc}  
 \omega - \epsilon_{1} +i \Gamma_{N} 
 -  { \left[t_{\sigma}^{2} /\left(\omega-\epsilon_{2}\right)\right] }
 & - \Gamma_{S} 
 \\
 - \Gamma_{S} 
 & \omega - \epsilon_{1} +i \Gamma_{N} 
 -  { \left[t^{2}_{\bar{\sigma}} / \left(\omega + \epsilon_{2}\right)\right] }
 \\ 
\end{array}\right) ^{-1}.
\label{Gr44}
\end{equation} 
Spectral function (local density of states) $\rho_{1 \sigma}(\omega)$ of QD$_{1}$ for each spin component $\sigma$ is given by standard formula
$\rho_{1 \sigma}(\omega) = -(1/ \pi) \textrm{Im} \left[ \check{G}^{11}_{1 \sigma}\left(\omega+i 0^{+}\right) \right]$.
The position of the Fermi level is $\omega=\mu_S$, which is located in the middle of the superconducting energy gap.
In the following, for a sake of simplicity, we also take that $\mu_S=\mu_N=0$.

\section*{Fano-like resonances in nanoscopic systems}

If low dimensional structures with discrete energy spectrum (such as, e.g., quantum dots) are coupled to reservoirs characterized by continuum of states, the energy levels of nanoobjects are broadened to form the Breit-Wigner (Lorentz) distribution with half-width controlled by QD-bath coupling strength (i.e., $\Gamma_N$).
Consequently, if subparts of a device are coupled to environment with different coupling strengths, broadening of particular energy levels differ significantly.
Quantum interference of electron waves resonantly transmitted through narrow (quasidiscrete levels) and those transmitted through broad levels give rise to asymmetric Fano-like profiles observed in density of states and differential conductivity.
For electrons whose energy tends to resonant energy from one side (e.g., $\omega \rightarrow \omega_{res}^{+}$), scattering does not change their phase, while for electrons reaching the resonant level from the other side ($\omega \rightarrow \omega_{res}^{-}$) phase is shifted by a factor of $\pi$ \cite{MiroshnichenkoRMP2010, ZitkoPRB2010}.
Therefore, in the Fano-like profiles a constructive enhancement (i.e., the $0$ phase shift) is accompanied by an
antiresonant deep (i.e., the $\pi$ phase shift).

In the original work \cite{UFano1935} and later review \cite{UFano1961}, U. Fano described the asymmetric lineshapes in the absorption spectra of noble gases observed previously by H. Beutler \cite{Beutler1935}. 
He introduced the formula for such profiles on a flat background. 
This function can be expressed as 
\begin{eqnarray}
\label{eq:Fano0}
F(\omega)= \frac{ {\left(q\Gamma_R/2 + \omega - \omega_{res}\right)^2}}{(\Gamma_R/2)^2+(\omega-\omega_{res})^2}, 
\end{eqnarray}
where $\omega_{res}$ and $\Gamma_{R}$ stand, respectively, for the energy of  {the} resonant level and  the effective broadening of  {the} resonant level, and $q$ represents  {a phenomenological} factor called an asymmetry parameter.

Because it will be very useful in the further discussion included in the next section (\emph{Profile analysis}), it is worth noticing a few properties of this function. 
Function (\ref{eq:Fano0}) has one local minimum and one maximum. 
 {Arguments of the extremes} are located  {at} $ {\omega_-}=\omega_{res}-q\Gamma_{R}/2$ and $ {\omega_+}=\omega_{res}+\Gamma_{R}/(2q)$, respectively. 
Due to the ideal anti-resonance, the minimum value of this function is equal to 0, while the maximum depends only on the asymmetry parameter and it is equal to $1+q^2$. 
Away from the resonant energy ( {i.e., for} $\omega \ll \omega_{res}$ or $\omega \gg \omega_{res}$)  {the} function  {reaches} value equals to 1. 
For asymmetry parameter close to unity and small broadening of the resonant energy level ($\Gamma_{R}$) the close proximity of  {the minimum and the maximum} forms a well-pronounced asymmetric profile.
Note that, for the asymmetry parameter approaching to zero, the Fano function corresponds to a symmetric deep, whereas for the parameter going to the infinity, the resonant feature resembles the Lorentz function.

In the nanoscopic systems, the broadening of the resonant level is dependent on the coupling to the continuum of states. 
Namely, in the case of the double quantum dot, it is proportional to the square of the interdot coupling. 
For  {the} strong interdot coupling, the broadening of the side level becomes comparable with the broadening of the central dot (cf., e.g., Refs. \cite{Gorski2018,Zienkiewicz2019} and references therein). 
In such a case, the interferometric structures evolve into molecular states.
Nevertheless, this issue is out of the scope of the present work, where we focus on the Fano-like features.

\subsubsection*{Fano-like resonances in a presence of superconducting electrode (symmetric case: $t_\uparrow=t_\downarrow \neq 0$)}

In a hybrid system, where single QD is coupled to SC reservoir,  due to proximity effects single particle QD's level evolves into two quasiparticle peaks representing so-called Andreev bound states (ABS). 
These states in noncorrelated regime emerge at $\omega=\pm E_{1}$, where quasiparticle energy levels are represented by $E_{1}=\sqrt{\epsilon_{1}^2+\Gamma_{S}^2}$, and they are weighted by a corresponding BCS coefficients 
$u^2 =\left(1 +\epsilon_{1}/E_{1}\right) / 2$ 
and 
$v^2 =\left(1 - \epsilon_{1}/E_{1}\right) / 2$. 
In a particular case of $\epsilon_{1}=0$, the Andreev states are symmetric Lorentzians separated by $\Gamma_{S}$. 
Density of states of the QD in such conditions (for $\epsilon_{1}=0$) can be expressed as
\begin{equation}
S(\omega)= \frac{1}{2} \left[ \frac{\Gamma_{N}}{(\omega-\Gamma_{S})^2+\Gamma_{N}^2}+ \frac{\Gamma_{N}}{(\omega+\Gamma_{S})^2+\Gamma_{N}^2} \right].
\label{eq:ABS0}
\end{equation}

If one QD (QD$_1$) is coupled to both metallic and superconducting electrodes and, additionally, side-coupled to  {the} second quantum dot (QD$_2$) with  {the} spin-independent coupling (i.e., the system shown in Fig.~\ref{fig:1.system}, but with $t_\uparrow=t_\downarrow$), the combined effect of the electron scattering on discrete level and the local pairing gives rise to two resonant features on background of ABS states [see Fig.~\ref{fig:23and4}(a)] \cite{BaranskiPRB2011,BaranskiPRB2012}. 
First one appears for energies close to energy level of  {the} side dot ($\omega \approx \epsilon_{2}$). 
Asymmetric lineshape of this resonance resembles the characteristic Fano-like shape.
Second feature emerges on opposite side of the Fermi level ($\omega \approx -\epsilon_{2}$). 
This resonance, however, differs significantly from the former one. 
First notable observation is a sharp spike apparent on one side of the resonance near $\omega\approx-\epsilon_{2}$. 
In terms of the Fano function such imbalance emerges for very large asymmetry parameters $q$. 
Second peculiar observation is that local minima near this resonance (in particular this for $\omega = -\epsilon_{2}$) have a finite value while the ordinary Fano function vanishes for $\omega=\omega_{res}-\left(q\Gamma_{R}\right)/2$.

\subsubsection*{Strongly asymmetric spin-polarized tunneling model ($t_\uparrow \neq 0$ and $t_\downarrow = 0$)} 
\label{sec:Fano-assymm}

\begin{figure}
\includegraphics[width=\size]{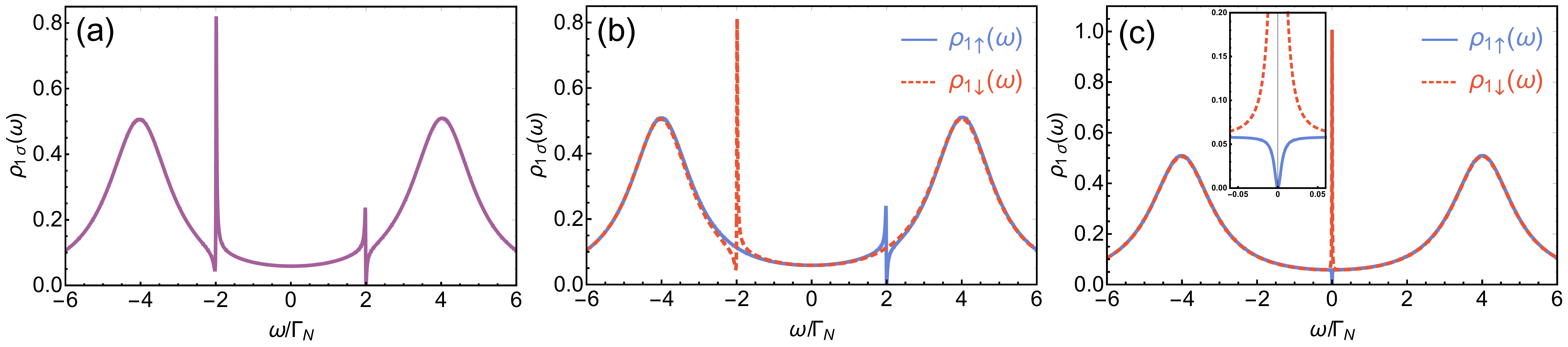}
\caption{%
(a) Spectral function $\rho_{1 \uparrow}(\omega)=\rho_{1 \downarrow}(\omega)$ of QD$_{1}$ for the spin-independent coupling with QD$_2$. 
The model parameters are: $t_{\uparrow}=t_{\downarrow}=0.3\Gamma_{N}$, $\Gamma_{S}=4\Gamma_{N}$, $\epsilon_{2}=2\Gamma_{N}$, $\epsilon_{1}=0$ (cf. also Ref. \cite{BaranskiPRB2012}).
(b) Spectral functions $\rho_{1 \uparrow}(\omega)$ (solid blue line) and $\rho_{1 \downarrow}(\omega)$ (dashed red line) of QD$_{1}$ for the spin-dependent coupling (the strongly asymmetric condition) with QD$_2$. 
The model parameters are:  $t_{\uparrow}=0.3\Gamma_{N}$, $t_{\downarrow}=0$, $\Gamma_{S}=4\Gamma_{N}$, $\epsilon_{2}=2\Gamma_{N}$, $\epsilon_{1}=0$.
(c) Spectral functions $\rho_{1 \uparrow}(\omega)$ (solid blue line) and $\rho_{1 \downarrow}(\omega)$ (dashed  red line) of QD$_{1}$ for the spin-dependent coupling (the strongly asymmetric condition) with QD$_2$.
In the inset the region near $\omega\approx 0$ is shown.
The model parameters are: $t_{\uparrow}=0.3\Gamma_{N}$, $t_{\downarrow}=0, \Gamma_{S}=4\Gamma_{N}$, $\epsilon_{2}=\epsilon_{1}=0$.
} 
\label{fig:23and4}
\end{figure}

To understand the origin of both resonances, we consider a (``toy'') model in which we completely suppress the interdot coupling for one spin component (e.g., for spin $\sigma=\downarrow$).
In the metal-hybrid structures two spin channels are independent. 
In a presence of superconducting electrode spin-$\uparrow$ and spin-$\downarrow$ electrons are bound into the local pairs. 
Thus, every physical process that involves one spin component affects also the other one.
By applying the strongly asymmetric condition (i.e., $t_{\downarrow}=0$), we effectively decompose the effect of  {the} direct interference, which occurs only for electrons coupled to  {the} side dot (i.e.,  {these} with $\sigma=\uparrow$) from the effect of bounding them into the local pairs  {[cf. Fig. \ref{fig:1.system}(b)]}. 
The  {latter} effect can be observed in the spectral function of electrons decoupled from the side dot ($\sigma=\downarrow$).
As spin-$\downarrow$ electrons are not directly scattered the resonant characteristics appearing in their spectral function originate solely from pairing with scattered electrons.

The spectral function of QD$_{1}$ for both directions of electron spin are shown in Fig.~\ref{fig:23and4}(b).
In such conditions, for directly scattered electrons ($\sigma=\uparrow$), we obtain only the resonant feature near $\omega\approx\epsilon_{2}$, while the second resonance disappears. 
Counter-wise, for opposite spin electrons (with $\sigma=\downarrow$) only the feature located near $\omega\approx-\epsilon_{2}$ remains.
It is worth noting that the shape of the resonant features remains unchanged, i.e.,  {the} shape of  {the} resonance near $\omega\approx\epsilon_{2}$ ($\omega\approx -\epsilon_{2}$) for  {the} symmetric case ($t_\uparrow=t_\downarrow \neq 0$) is identical as  {the} resonant feature that remains in $\rho_{1\uparrow}(\omega)$ ($\rho_{1\downarrow}(\omega)$, respectively) for  {the} perfectly-polarized interdot coupling ($t_\uparrow\neq t_\downarrow = 0$), cf. Fig. \ref{fig:23and4}(a) and Fig. \ref{fig:23and4}(b). 
Thus, this indicate that the resonance near $\omega\approx\epsilon_{2}$ originates purely from the direct scattering of electrons on a side level, while the resonant characteristic near $\omega\approx-\epsilon_{2}$ is merely a response to  {the} pairing of a given electron with its scattered partner.

In other words, one spin component is directly scattered on side structure while the other one  {``feels''} the scattering only by bounding into a local pair with directly scattered one. 
Such the conclusion is also visible if we compare the analytic formula for single particle Green's functions for each spin component at QD$_1$. 
These functions for perfectly spin-polarized tunneling ($t_{\downarrow}=0$) are represented by
\begin{eqnarray}
\label{eq:Gupdown}
\langle \langle \hat{d}_{1 \uparrow} \hat{d}^{\dagger}_{1 \uparrow} \rangle\rangle  
  =   \left(\omega- \epsilon_{1}+ i\Gamma_{N} -  {K_{\uparrow}(\omega)}- \frac{\Gamma_{S}^2}{\omega+\epsilon_{1}+i\Gamma_{N}}\right)^{-1}, 
\quad
\langle \langle \hat{d}_{1 \downarrow} \hat{d}^{\dagger}_{1 \downarrow} \rangle\rangle = 
 \left(\omega- \epsilon_{1}+ i\Gamma_{N} - \frac{\Gamma_{S}^2}{\omega+\epsilon_{1}-  {K_{\uparrow}^{*}(\omega)} +i\Gamma_{N}}\right)^{-1},  
\end{eqnarray}
where  {$K_{\sigma}(\omega)=t_{\sigma}^2/(\omega-\epsilon_{2})$ and $K^{*}_{\sigma}(\omega)=t_{\sigma}^2/(\omega+\epsilon_{2})$ are parts responsible for the scattering [cf. also Eq. (\ref{Gr44})].}
 {The scattering} enters the spin-$\uparrow$ propagator totally independently of pairing $\Gamma_S$, cf. the fourth term  {$K_{\uparrow}(\omega)$} in  {the} expression for $\langle \langle \hat{d}_{1 \uparrow} \hat{d}^{\dagger}_{1 \uparrow} \rangle\rangle$.
For spin-$\downarrow$ electrons, the response for scattering of spin-$\uparrow$ electrons is provided solely by local pairing $\Gamma_{S}$, cf. the last term  {with $K^{*}_{\uparrow}(\omega)$ in the denominator} in $\langle \langle \hat{d}_{1 \downarrow} \hat{d}^{\dagger}_{1 \downarrow} \rangle\rangle$.

{It should be noticed that, in the case of the arbitrary tunneling (i.e., any $t_{\uparrow}\neq 0$, $t_{\downarrow} \neq 0 $), the self-energies of both types of electrons (i.e., with each spin direction $\sigma=\uparrow,\downarrow$) are composed of the part responsible for the direct scattering [connected with $K_{\sigma}(\omega)\propto t_{\sigma}^{2}$ term] as well as the part related to the pairing ($\propto \Gamma_S^2$) with scattered electrons [i.e., the pairing with the convoluted scattering, associated with $K^{*}_{\bar{\sigma}}(\omega)\propto t_{\bar{\sigma}}^{2}$ term], namely:		
\begin{eqnarray*}
\label{eq:Gboth}
\langle \langle \hat{d}_{1 \sigma} \hat{d}^{\dagger}_{1 \sigma} \rangle\rangle  
=   \left(\omega- \epsilon_{1}+ i\Gamma_{N} - K_{\sigma}(\omega)- \frac{\Gamma_{S}^2}{\omega+\epsilon_{1}- K^{*}_{\bar{\sigma}}(\omega) +i\Gamma_{N}}\right)^{-1}. 
\end{eqnarray*}	
Therefore, for the nonpolarized case, we observe both the ordinary Fano feature near $\omega\approx\epsilon_2$ and the ``anomalous'' Fano resonance near $\omega\approx-\epsilon_2$ for electrons with spin-$\uparrow$ as well as for electrons with spin-$\downarrow$.
For $t_{\downarrow}=0$, the above equations for $\sigma={\uparrow,\downarrow}$ reduces to Eq. (\ref{eq:Gupdown}) and each resonant feature occurs in different spin channel.}

The local density of states at QD$_1$ for each spin component is given by  {the} imaginary part of  {an} adequate Green function 
$\rho_{1\sigma}(\omega)=-(1 / \pi) \textrm{Im} \left[ \check{G}^{11}_{1 \sigma}\left(\omega+i 0^{+}\right) \right]=-(1/\pi) \textrm{Im} \langle \langle \hat{d}_{1 \sigma} \hat{d}^{\dagger}_{1 \sigma} \rangle\rangle$. 
In the case of $t_{\downarrow}=0$ and the symmetric Andreev states (i.e., $\epsilon_{1}=0$), Eq. (\ref{eq:Gupdown}) yields the following expressions for LDOS of each spin
\small
\begin{eqnarray}
\label{eq:rho}
\rho_{1\uparrow}(\omega)
 =  \frac{\frac{1}{\pi}\Gamma_{N} \left(\frac{\Gamma_{S}^2}{\Gamma_{N}^2+\omega^2}+1\right)}{\left(\omega+\frac{t_{\uparrow}^2}{\epsilon_{2}-\omega}-\frac{\omega \Gamma_{S}^2 }{\Gamma_{N}^2+\omega^2}\right)^2+\left(\frac{\Gamma_{N} \Gamma_{S}^2}{\Gamma_{N}^2+\omega^2}+\Gamma_{N}\right)^2}, \quad \textrm{and} \quad
\rho_{1\downarrow}(\omega) 
 = \frac{\frac{1}{\pi}\Gamma_{N} \left(\frac{\Gamma_{S}^2}{[f_{r}(\omega)]^2+\Gamma_{N}^2}+1\right)}{\left(\frac{\Gamma_{N} \Gamma_{S}^2}{[f_{r}(\omega)]^2+\Gamma_{N}^2}+\Gamma_{N}\right)^2+\left(\omega-\frac{\Gamma_{S}^2 f_{r}(\omega)}{[f_{r}(\omega)]^2+\Gamma_{N}^2}\right)^2},
\end{eqnarray}
\normalsize
where $f_{r}(\omega)=\omega-t_{\uparrow}/\left(\omega+\epsilon_{2}\right)$.

From the above equations, it is difficult to see if the shape around the resonant energies $\omega \approx \pm\epsilon_2$ can be described as the Fano-like shape. 
Moreover, for $\epsilon_2=0$ the sharp resonant peak at $\omega\approx - \epsilon_2$ evolves into a symmetric Lorenzian [cf. Fig. \ref{fig:23and4}(c)]. 
In the next section, we present analysis of resonant features in both spin channels for $t_{\downarrow}=0$.
For a sake of simplicity we focus on the case of $\epsilon_{1}=0$, which is studied further in this work.

\section*{Profiles analysis (asymmetric case: $t_{\downarrow}=0$)}

The Fano resonances were successfully used as a probe for electron phase coherence in quantum dots \cite{Clerk2001}. 
It was shown that dephasing time can be determined from the asymmetry parameter ($q$) of measured profiles. 
This issue was particularly relevant to take the meaning of Fano profiles appearing in single-electron transistor \cite{GoresPRB2000}). 
The influence of such dephasing on Fano resonances was also analyzed by one of us \cite{BaranskiPRB2012}. 
Thus, a proper evaluation of the asymmetry parameter for a given profile turn out to be a relevant issue. 
In the case of resonances that appear on non-flat backgrounds, straight fitting of the regular Fano function may produce highly inaccurate values. 
The problem becomes even more complicated if a given shape deviates from the regular Fano profile. 
In here analyzed system the resonances near $\omega\approx -\epsilon_{2}$ exhibit features that do not match the ordinary Fano shape.

In this section, we will analyze the obtained resonant lines to compare them with the Fano profiles and indicate to what extent a given profile can be approximated by the Fano function [cf. Eq. (\ref{eq:Fano0})].  
We develop a feasible procedure of fitting the Fano parameters to the assumed form. 
In the case of resonances that deviate from the ordinary Fano shape, we take into account and estimate the correction factor $\phi_{0}$.

In general, the Fano-like resonances can be represented as a function $\alpha F(\omega)$, where $F(\omega)$ [given by Eq. (\ref{eq:Fano0})] depends on the parameters $q$, $\Gamma_{R}$, $\omega_{res}$ and constant $\alpha$ is a flat background (in the original works \cite{UFano1935,UFano1961} $\alpha=1$ and the resonance appears on a flat singular background). 
In our case, the resonant features emerge on the background of quasiparticle Andreev states described by Eq. (\ref{eq:ABS0}).
Therefore, we assume that one can approximate the density of states $\rho_{1\sigma}(\omega)$ by a product of the ordinary Fano curve and a background composed of the Andreev states, i.e., by $R_{\sigma}=F_\sigma(\omega)S(\omega)$, where $S(\omega)$ is given by (\ref{eq:ABS0}). 
If so, it should be possible to find the relations between the model parameters and the parameters used in the function $F_\sigma(\omega)$: $q_\sigma$, $\Gamma_{R,\sigma}$, and $\omega_{res,\sigma}$ in such way that the constructed function $F_\sigma(\omega)S(\omega)$ will reproduce the density of states $\rho_{1\sigma}(\omega)$ with high accuracy.

\subsection*{The resonance for directly scattered electrons (near $\omega\approx \epsilon_2$)}

\begin{figure}
\includegraphics[width=\size]{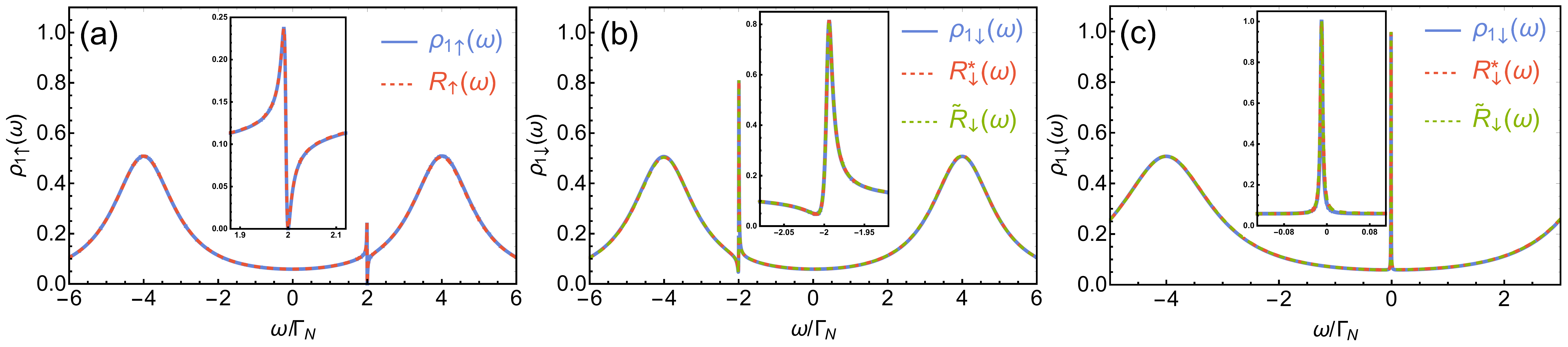}
\caption{
(a) Convergence of assumed form $R_{\uparrow}(\omega)$ (dashed red line) and exact function $\rho_{1\uparrow}(\omega)$ (solid blue line) obtained for following model parameters: 
$\Gamma_{S}=4\Gamma_{N}$, $t_{\uparrow}=0.2\Gamma_{N}$, $\epsilon_{2}=2\Gamma_{N}$.
In the inset the region near $\omega\approx\epsilon_2$ is enlarged.
(b) Convergence of assumed forms $R^{*}_{\downarrow}(\omega)$ (dashed red line), $\tilde{R}_{\downarrow}$ (dotted green line) and exact function $\rho_{1\uparrow}(\omega)$ (solid blue line) obtained for following model parameters: 
$\Gamma_{S}=4\Gamma_{N}$, $t_{\uparrow}=0.2\Gamma_{N}$, $\epsilon_{2}=2\Gamma_{N}$.
(c) Convergence of assumed forms $R^{*}_{\downarrow}(\omega)$ (dashed red line), $\tilde{R}_{\downarrow}$ (dotted green line) and exact function $\rho_{1\uparrow}(\omega)$ (solid blue line) obtained for following model parameters: 
$\Gamma_{S}=4\Gamma_{N}$, $t_{\uparrow}=0.2\Gamma_{N}$, $\epsilon_{2}=0.001\Gamma_{N}$.}
\label{fig:56and7}
\end{figure}

We will start with an analysis of the LDOS $\rho_{1\uparrow}(\omega)$ for electrons directly scattered on side dot QD$_2$. 
For weak scattering (i.e., for $t^{2}_\uparrow \ll \Gamma_{N}^2$), the resonant feature in exact function $\rho_{1\uparrow}(\omega)$ is represented by a sharp deep-spike characteristic. 
The bare Andreev states are represented by smooth Lorentzians with the half-width controlled by $\Gamma_{N}$. 
In such a case, one can assume that arguments $\omega$ for which the $F_{\uparrow}(\omega)$ takes the minimum ($\omega=\omega_-$) and maximum ($\omega=\omega_+$) should be very close to local extremes of product function $R_{\uparrow}(\omega)=F_{\uparrow}(\omega)S(\omega)$, where
$F_\uparrow(\omega)$ has a form of Eq. (\ref{eq:Fano0}) with $q_\uparrow$, $\Gamma_{R,\uparrow}$, and $\omega_{res,\uparrow}$ parameters. 
On the other hand, maximum value of the Fano function $F_{\uparrow}(\omega_+)$ is dependent only on asymmetry parameter $F_{\uparrow}(\omega_+)=1+q_{\uparrow}^{2}$ ($\omega_+$ is location of the maximum).
Therefore,  {an} expression for asymmetry parameter $q_{\uparrow}$ can be obtained from a maximum of the exact function. 
Assuming that $\omega_{+}$ is an argument of the local maximum of $\rho_{1\uparrow}(\omega)$ (around $\omega\approx\epsilon_{2}$) we have 
\begin{equation}
\rho_{1\uparrow}(\omega_{+})=(1+q_{\uparrow}^2)S(\omega_{+}).
\label{eq:qup}
\end{equation}
The sign of asymmetry parameter $q_{\uparrow}$ is governed by a position of $\epsilon_{2}$ 
(i.e., for $\epsilon_{2}>0 $ one gets $q_{\uparrow}<0$ and for $\epsilon_{2}<0$ one has $q_{\uparrow}>0$). 
The other two parameters $\omega_{res,\uparrow}$ and $\Gamma_{R,\uparrow}$ can be found by comparison of positions of local minimum and maximum of $\rho_{1\uparrow}(\omega)$ and $F_{\uparrow}(\omega)$. 
Arguments $\omega$ for which the Fano function takes the minimum is given by $\omega_{-}=\omega_{res,\uparrow}-q_\uparrow \Gamma_{R,\uparrow}/2$, while for the maximum $\omega_{+}=\omega_{res,\uparrow}+\Gamma_{R,\uparrow}/(2 q_\uparrow) $. 
This yields 
\begin{equation}
\omega_{res,\uparrow}=\frac{\omega_{-}+q_\uparrow^2\omega_{+} }{1+q_\uparrow^2}, \qquad
\Gamma_{R,\uparrow}=\frac{2 q_\uparrow (\omega_{+}-\omega_{-})}{1+q_\uparrow ^2},
\end{equation}
where $q_{\uparrow}$ is an asymmetry parameter estimated previously from Eq. (\ref{eq:qup}). 
In Fig. \ref{fig:56and7}(a) we examine the convergence of the obtained function with the exact prototype. 
One can note that for spin-$\uparrow$ electrons  {the} product function with $q_{\uparrow},\Gamma_{R,\uparrow}$ and $\omega_{res,\uparrow}$ estimated by the above procedure reproduces original $\rho_{1\uparrow}(\omega)$ with very high accuracy.
Asymmetry parameter $q_\uparrow$ for this fit of the Fano resonant feature in $\rho_{1\uparrow}(\omega)$ as a function of $\epsilon_2$ is shown in Fig. \ref{fig:8and9}(b) as dotted green line.

\subsection*{The resonance for indirectly scattered electrons (near $\omega\approx -\epsilon_2$)}

\subsubsection*{Approximation with the regular Fano function}

Situation is more complicated for opposite spin electrons, i.e., for spin-$\downarrow$ electrons. 
The minimum value of the exact function is finite, while minimal value for the ordinary Fano resonance is equal to $0$. 
This means that product of the ordinary Fano resonant curve and arbitrary background [in particular, also function  {$R_{\downarrow}(\omega)=F_{\downarrow}(\omega)S(\omega)$}] will never reproduce the original function $\rho_{1\downarrow}(\omega)$ for spin-$\downarrow$ electrons accurately. 
However, we can still assume that spectral function can be approximated by  {the} Fano function with small correction $\phi_{0}$. 
 {Let us} assume that  {the} exact function can be approximated by a product of $S(\omega)$ and
\begin{eqnarray}
F^{*}_{\downarrow}(\omega)= F_{\downarrow}(\omega) +\phi_{0}, 
\end{eqnarray}
where $\phi_{0}$ represents a (small) deviation (correction) dependent only on model parameters (and it is independent of $\omega$) and
$F_\downarrow(\omega)$ has a form of Eq. (\ref{eq:Fano0}) with $q_\downarrow$, $\Gamma_{R,\downarrow}$, and $\omega_{res,\downarrow}$ parameters. 
One should note that if correction parameter $\phi_{0}$ will be small enough (say less than $1$) then one can state that the  {resonant} shape can be approximated by the Fano function with high accuracy. 
Thus, auxiliary parameter $\phi_{0}$ can be considered as a measure to what extent one can fit the original Fano shape into a given resonant feature. 
Therefore,  {a} high  value of correction $\phi_{0}$ indicate that the Fano function may not be adequate for fitting to  {the} exact function.
Introduction of $\phi_{0}$ slightly rearranges the background of the ordinary Fano function, i.e., away from the resonant energy $F^*_{\downarrow}(\omega)$ reaches $1+\phi_{0}$ instead of just $1$.
This would cause problems in achieving acceptable convergence of assumed form and the exact function away from the resonant feature.  
To neutralize this obstacle we need to normalize our assumption. 
Therefore, instead of $F^*_{\downarrow}(\omega)S(\omega)$, we assume  that $\rho_{1\downarrow}(\omega)$ can be approximated by
\begin{equation}
\label{eq:rstardown}
R^*_\downarrow(\omega)=\frac{F^*_{\downarrow}(\omega)S(\omega)}{1+\phi_{0}}=\frac{F_{\downarrow}(\omega) +\phi_{0}}{1+\phi_{0}}S(\omega).
\end{equation}

With this modification we find that  {a} value in  {the} minimum of $R^*_\downarrow(\omega)$ is dependent solely on $\phi_0$, while value in maximum of $R^*_\downarrow(\omega)$ depends on $\phi_{0}$ and $q_\downarrow$. 
Positions  {of $\omega_{\mp}$}, for which $R^*_\downarrow(\omega)$ gets its minimal and maximal values, remain unchanged  {[i.e., we assume that they are the same as those for $F_{\downarrow}(\omega)$]}.
 {Extreme arguments $\omega_{-}$ and $\omega_{+}$ 	can be calculated numerically (in experimental realizations these values can be directly read from the data).}
Therefore,   {we should add just one step in our procedure}. 
First, we get $\phi_{0}$ comparing local minima of $\rho_{1\downarrow}(\omega)$ 
and assumed form (\ref{eq:rstardown}) of $R^*_\downarrow(\omega)$.
 {Equation (\ref{eq:rstardown}) with $F_{\downarrow}(\omega)$ vanishing for $\omega = \omega_{-}$ yields $\phi_0= \rho_{1\downarrow}(\omega_-)/\left[\rho_{1\downarrow}(\omega_-) - S(\omega_-)\right]$.} 
 {Then,} we find asymmetry parameter $q_\downarrow$ by comparing the  {values at maxima of $\rho_{1\downarrow}(\omega_{+})$ and $R^*_\downarrow(\omega_{+})$}.
 {This gives $q_{\downarrow}^2=\left[\rho_{1 \downarrow}(\omega_{+})-S(\omega_{+})](1+\phi_{0})\right]/S(\omega_{+})$ [using the property of the Fano function that $F_{\downarrow}(\omega_+)=1+q_{\downarrow}^{2}$].} 
By means of acquired $\phi_0$ and $q_\downarrow$, we obtain resonant energy $\omega_{res,_\downarrow}$ and broadening $\Gamma_{R,_\downarrow}$.  
Using the above procedure, we find that  {the} exact function $\rho_{1\downarrow}(\omega)$ can be approximated with good accuracy by $R_\downarrow(\omega)$, [cf.  dashed  red line and  solid blue line in Figs. \ref{fig:56and7}(b) and \ref{fig:56and7}(c)].
However, good convergence with  {a} small correction is achieved only for resonant energies close to the center of Andreev states ($\epsilon_{2}\approx \pm\Gamma_{S}$) (see solid blue line in Fig. \ref{fig:8and9}(a)).
For resonant energies close to  {the} Fermi level, spectral function  {$\rho_{1\downarrow}(\omega)$} is represented by almost  {a} symmetric peak [cf. Fig. \ref{fig:23and4}(c)]. 
For the original Fano curve (\ref{eq:Fano0}) such a situation is met when  {the} asymmetry parameter approach infinity ($q\rightarrow\infty$). 
On the other hand, a very large asymmetry parameter implies huge values for the maximum of the Fano function  {$F_{\downarrow}(\omega)$}. 
For the product to remain finite, the correction must grow together with the asymmetry parameter.  
Thus, in  assumed form  {$R^{*}_{\downarrow}(\omega)$} of  {$\rho_{1\downarrow}(\omega)$ for} $\epsilon_{2}\approx 0$, ``correction'' $\phi_0$ (as well as asymmetry parameter $q_\downarrow$) become enormously large (cf. Fig. \ref{fig:8and9}(a); they tend to infinity if $\epsilon_{2} \rightarrow 0$). 
A correction value is also inadequate if  {the} resonant energy is located far outside  {the} Andreev states (i.e., $|\epsilon_{2}| \gg \Gamma_{S}$) .
These make the statement about such resonances as the Fano-like one somewhat exaggerated.
Asymmetry parameter $q_\downarrow$ of $F_{\downarrow}(\omega)$ for this fit of the resonant feature in $\rho_{1\downarrow}(\omega)$ as a function of $\epsilon_2$ is shown in Fig. \ref{fig:8and9}(b) as  solid blue line.

\subsubsection*{Approximation with the inverse of the Fano function}

\begin{figure}
\centering
\includegraphics[width=\size]{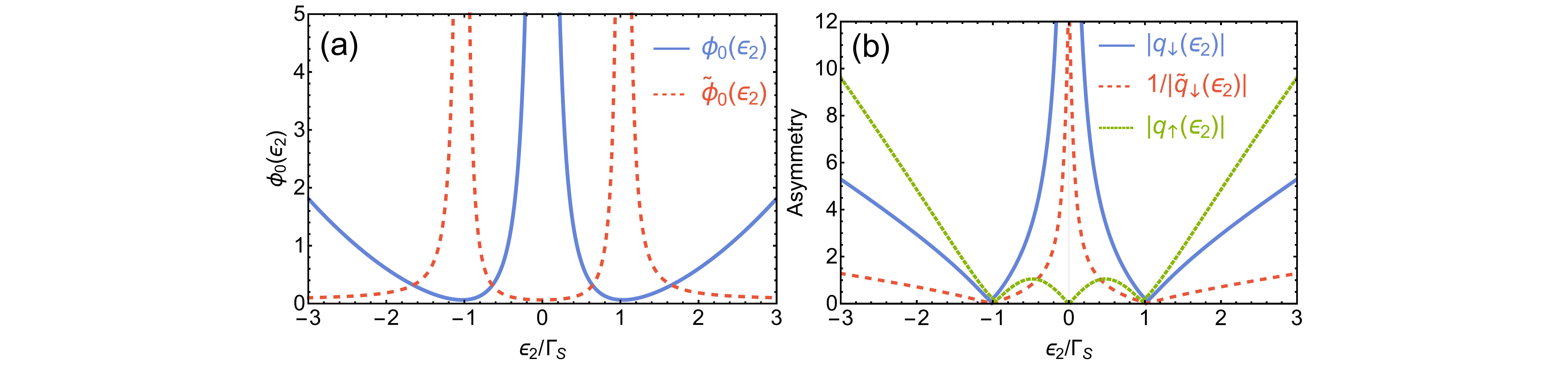}
\caption{%
(a) Corrections values ($\phi_{0}$ and $\tilde{\phi}_{0}$) as a function of energy of  {the} side dot $\epsilon_{2}$ for both assumptions for anomalous Fano resonant features of indirectly scattered electrons. 
Solid blue line refers to $\phi_{0}$ in the assumed form of Eq. (\ref{eq:rstardown}), while dashed red  line is for $\tilde{\phi}_{0}$ as assumed in Eq. (\ref{eq:rtilde}).
(b) Absolute values of asymmetry parameters as a function of  side-dot energy $\epsilon_2$:  
$q_{\downarrow}$ used in $F_\downarrow^*(\omega)$ of Eq. (\ref{eq:rstardown}) (solid blue line), 
and 
$1/\tilde{q}_{\downarrow}$ used in $\tilde{F}_\downarrow(\omega)$ of Eq. (\ref{eq:rtilde}) (dashed red line),
and 
$q_{\uparrow}$ used in $F_\uparrow(\omega)$ (dotted green line).
The data on both panels are obtained for $\Gamma_{S}=4\Gamma_{N}$.}
\label{fig:8and9}
\end{figure}

 {To underline the fact that the ordinary Fano function is not the best way to approximate the resonant feature near $\omega\approx-\epsilon_2$ (at least at same range of the model parameters),  we will try to fit another asymmetric function with a well-defined asymmetry parameter and compare the result with fitting of the ordinary Fano function.}
The problem of  {an} inadequate correction for a wide spectrum of $\epsilon_{2}$ can be reduced by  {a} slight rearrangement of the assumption. 
The resonant feature in $\rho_{1\downarrow}(\omega)$ is composed of a finite deep accompanied by an over-sized peak.
We noticed that if we add a small parameter to the regular Fano function, and then we take the inverse of that structure, the resulting function should have similar features. 
Thus, we propose to approximate  {the} original function by a product of  {the} Andreev states  {[i.e., $S(\omega)$]} and an inversion of the Fano function. 
Taking into account the normalization as previously, our assumption should be in the following form
\begin{equation}
\tilde{R}_\downarrow(\omega)=\frac{1+\tilde{\phi}_{0}}{\tilde{F}_\downarrow(\omega) +\tilde{\phi}_{0}}S(\omega), 
\label{eq:rtilde}
\end{equation}
where $\tilde{F}_\downarrow(\omega)$ has a form of Eq. (\ref{eq:Fano0}) with $\tilde{q}_\downarrow$, $\tilde{\Gamma}_{R,\downarrow}$, and $\tilde{\omega}_{res,\downarrow}$. 
Using the procedure similar to the previous one, we determine the corresponding parameters and compared the obtained result with the exact function  {$\rho_{1\downarrow}(\omega)$}.
 {Here, maximum (minimum) of $\tilde{F}_{\downarrow}(\omega)$ at $\omega_{+}$ ($\omega_{-}$, respectively) corresponds to minimum (maximum) of $\rho_{1\downarrow}(\omega)$.
Thus, using the properties of the Fano curve $\tilde{F}_{\downarrow}(\omega)$ in the similar manner as previously, one gets that $\tilde{\phi}_0= S(\omega_-)/\left[\rho_{1\downarrow}(\omega_-) - S(\omega_-)\right]$ and $\tilde{q}_{\downarrow}=(1+\tilde{\phi}_{0})\left[S(\omega_{+})-\rho_{1\downarrow}(\omega_+)\right]/\rho_{1\downarrow}(\omega_+)$.
} 
We found that  {the} new assumption reproduces the original function as accurately as the previous one  {[cf. Figs. \ref{fig:56and7}(b) and \ref{fig:56and7}(c)]}. 
The advantage  {of such a fit is } that correction $\tilde{\phi}_{0}$ for such assumption is considerably smaller for  {a} wide range of $\epsilon_{2}$ excluding $\epsilon_{2} \approx \pm \Gamma_{S}$, where  {it expands} to infinity (Fig. \ref{fig:8and9}(a)). 
If the resonant energy is close to $\Gamma_{S}$,  {the} spectral function can be approximated by the regular Fano shape with a small correction using Eq. (\ref{eq:rstardown})  {again,} as described in previous section. 
 {The} inverse of asymmetry parameter $\tilde{q}_\downarrow$ of $\tilde{F}_{\downarrow}(\omega)$ for this fit of the resonant feature in $\rho_{1\downarrow}(\omega)$ as a function of $\epsilon_2$ is shown in Fig. \ref{fig:8and9}(b) as  dotted red line.
Note, that in this case, $1/\tilde{q}_{\downarrow}$ (rather than $\tilde{q}_{\downarrow}$) is a measure of  {the} asymmetry comparable to the ordinary Fano asymmetry parameter (as for $1/\tilde{q}_{\downarrow} \rightarrow 0$ resonant feature is represented by a symmetric deep and for $1/\tilde{q}_{\downarrow} \rightarrow \infty$ by the Lorentz distribution).

One should note that, in  {contrast} to the Fano function, which originates from a rigorous examination of transmission rates in noble gases \cite{FANO1968}, function (\ref{eq:rtilde}) is a hypothetical (semiempirical) function that can be fitted into ``anomalous Fano'' curves more accurately at wide range of parameters. 
Nevertheless, using such a function one can estimate the parameter $\tilde{q}_{\downarrow}$, which is responsible for a measure of an asymmetry of the resonant feature (and thus, it is fragile for decoherence).

Concluding, the LDOS function for indirectly scattered electrons can be treated as a normalized product of the Andreev states and (i) Fano resonance or (ii) inverse Fano, both with a small correction. 
The first approach reproduces well the original function only for resonant energies close to $\Gamma_{S}$, i.e., it describes a case when the resonant energy coincide with the Andreev states. 
The second approach works well for resonant energies much smaller and much larger than $\Gamma_{S}$.
To have a full insight into the behavior of LDOS function, it would be useful to combine these two approaches or determine the ratio $\epsilon_2/\Gamma_{S}$ and then use appropriate product.

\section*{Effects of correlations between electrons on the quantum dots}

In nanoscopic systems, the Coulomb repulsion between electrons often plays an important role, therefore, in this section, we briefly discuss the interplay of correlations effects with the analyzed features. 
In the model, QD$_{2}$ is not directly connected to any external reservoir and on-site interactions on QD$_2$ (i.e., $U_{2}$) lead only to appearance of an additional narrow state in the spectrum of QD$_{2}$ located at $\omega=\epsilon_{2}+U_{2}$. 
Consequently, the influence of such interactions on spectrum of QD$_1$ is straightforward.  
For the perfectly polarized case (i.e., $t_{\downarrow}=0$), two resonant Fano-like features emerge in $\rho_{1\uparrow}(\omega)$ at energies $\omega=\epsilon_{2}$ and $\omega=\epsilon_{2}+U_{2}$, as a consequence of direct scattering. 
In the spectrum of opposite spin electrons  {[i.e., $\rho_{1\downarrow}(\omega)$]}, two anomalous resonances are formed at the opposite side of the Fermi level, i.e.,  at $\omega=-\epsilon_{2}$ and $\omega=-(\epsilon_{2}+U_{2})$. 
In the case of $t_{\uparrow}=t_{\downarrow}$, all four features emerge (as shown  {in detail} in Ref. \cite{BaranskiPRB2011}).

\begin{figure}
	\centering
	\includegraphics[width=\size]{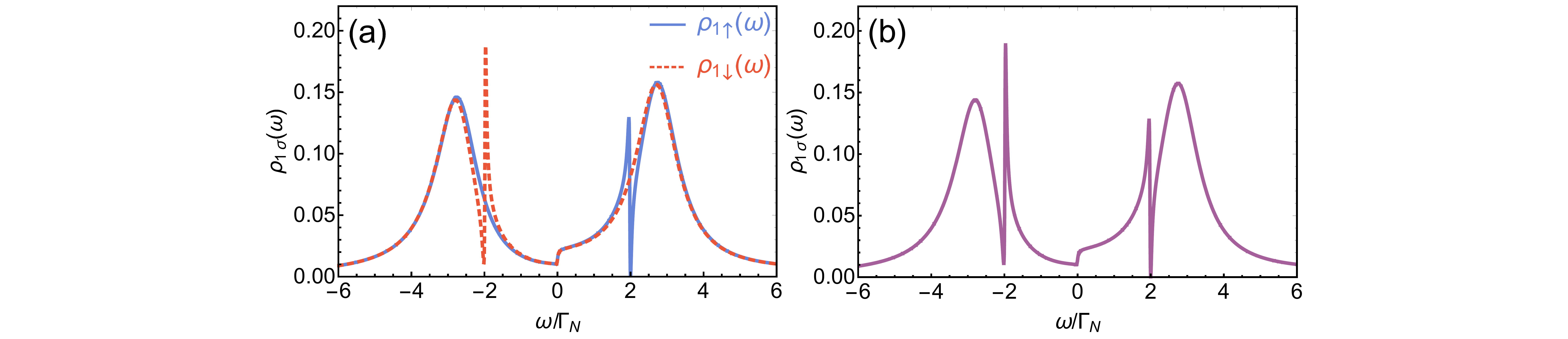}
	\caption{ {(a)} Spectral functions  $\rho_{1 \uparrow}(\omega)$ (solid blue line) and $\rho_{1 \downarrow}(\omega)$ (dashed red line) of QD$_1$ in the strongly correlated regime $U_{1}=15\Gamma_{N}$ (and $U_2=0$) for  {the} perfectly spin-polarized interdot hopping (i.e., $t_{\uparrow}=0.3\Gamma_{N}$, $t_{\downarrow}=0$), energy of the side dot $\epsilon_{2}=2\Gamma_{N}$, strong hybridization to SC electrode $\Gamma_{S}=4\Gamma_{N}$, energy of the interfacial dot $\epsilon_1=0$, and low temperature $k_{B}T=0.01\Gamma_{N}$.
		 {(b) Spectral function $\rho_{1\uparrow}(\omega)=\rho_{1\downarrow}(\omega)$ for the nonpolarized model with $t_{\uparrow}=t_{\downarrow}=0.3\Gamma_{N}$. Other parameters are the same as on panel (a).}
	} 
	\label{fig:corr1}
\end{figure}

To account for the correlations on QD$_{1}$, we adopt procedure used previously in Ref. \cite{BaranskiPRB2011}.
In the presence of correlations, the matrix of Green's functions can be represented by 
\begin{equation}
\check{\mathcal{G}}^{-1}_{1 \sigma}(\omega) =
\left( \begin{array}{cc}  
\omega - \epsilon_{1} -\frac{t_{\sigma}^2}{\omega-\epsilon_{2}} & - \Gamma_{S} \\
- \Gamma_{S} & \omega - \epsilon_{1} -\frac{t^2_{\bar{\sigma}}}{\omega + \epsilon_{2}}
\\ 
\end{array}\right) - \left ( \begin{array}{cc}  
\Sigma_{N,\sigma}(\omega) & 0\\
0 & -\Sigma^{*}_{N,\sigma}(-\omega)
\\ 
\end{array}\right),
\label{eq:Gcorr}
\end{equation} 
where self energy $\Sigma_{N}(\omega)$ is approximated using the decoupling scheme, which approximates higher order Green's functions and reduces them to lower order once; details are given in Ref.\cite{BaranskiPRB2011}. 
It yields:
\begin{equation}
\Sigma_{N,\sigma}(\omega)\simeq \omega-\epsilon_{1}-\frac{(\omega-\epsilon_{1}-\Sigma_{0})[(\omega-\epsilon_{1}-\Sigma_{0})-U_{1}-\Sigma_{3}]+U_{1}\Sigma_1}{\omega-\epsilon_{1}-\Sigma_{0}-\Sigma_{3}-(1-\langle n_{\bar{\sigma}} \rangle)U_{1}},
\end{equation}
where $\Sigma_{\eta=1,3}$ are given by
\begin{equation}
\Sigma_{\eta} = \sum_{\bf{k}}|V_{\bf{k} N}|^2\left(\frac{1}{\omega+\xi_{k N}-2\epsilon_{1}-U_{1}} + \frac{1}{\omega-\xi_{k N}} \right)[f(\omega)]^{3-\eta/2}
\end{equation}
with $f(\omega)=1/ \left[1+\exp{\left(\omega/k_{B}T\right)}\right]$ being the Fermi distribution at temperature $T$, $\langle n_{\bar{\sigma}} \rangle$ denotes an average occupancy of QD$_1$ with spin-$\bar{\sigma}$ electrons (calculated self-consistently),  and $\Sigma_{0}=i \Gamma_{N}/2$.
In this section, we will investigate the spin-dependent energy spectrum in the perfectly polarized case, i.e., $t_{\downarrow}=0$. 
The symmetric spin interdot coupling case was described in Ref. \cite{BaranskiPRB2011}. 
We inspect two cases:
(i) the strongly proximized case, where the hybridization to superconducting electrode is considerably larger than coupling to metallic one, i.e, $\Gamma_{S}=4\Gamma_{N}$ (used also in the previous sections of the present work) and 
(ii) the case with comparable hybridizations, namely  $\Gamma_{S} = \Gamma_{N}$. 
For the former conditions, it is possible to inspect the interplay between the Kondo physics and the resonant features originating from electron scattering.

\begin{figure}
	\centering
	\includegraphics[width=\size]{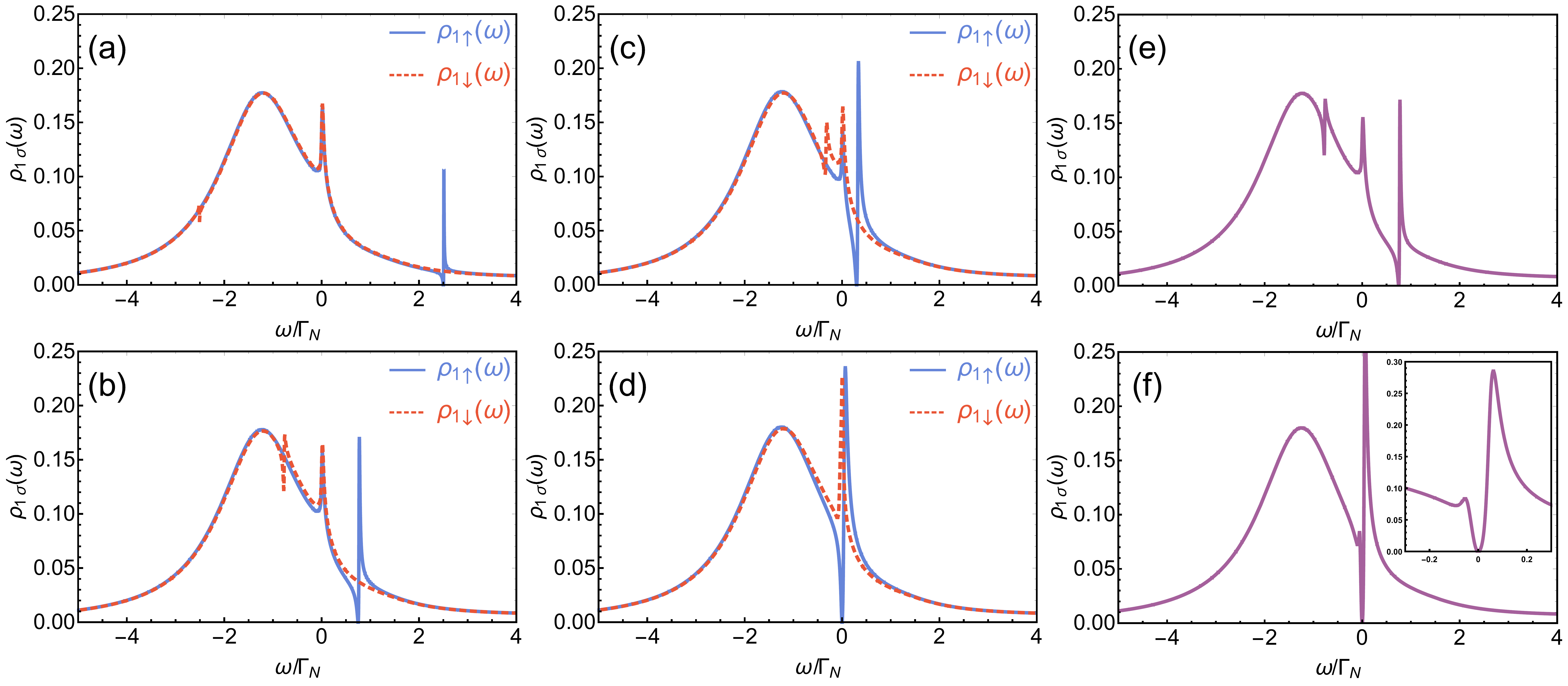}
	\caption{%
	 {(a)--(d)} Spectral functions  $\rho_{1 \uparrow}(\omega)$ ( {solid} blue line) and $\rho_{1 \downarrow}(\omega)$ ( {dashed} red line) of QD$_1$ obtained for  {the} spin-polarized interdot hopping  
	(i.e., $t_{\uparrow}=0.3\Gamma_{N}$, $t_{\downarrow}=0$) 
	and in the Kondo regime. 
	The model parameters are: 
	$U_{1}=15\Gamma_{N}$, $U_2=0$, 
	$\Gamma_{S}=\Gamma_{N}$, $\epsilon_1=-1.5\Gamma_N$, 
	and $k_{B}T=0.01\Gamma_{N}$. 
	The side-dot energy for each panel is: 
	(a) $\epsilon_{2}=2.5\Gamma_{N}$, 
	(b) $\epsilon_{2}=0.75\Gamma_{N}$, 
	(c) $\epsilon_{2}=0.3\Gamma_{N}$, and 
	(d) $\epsilon_{2}=0$.
	 {(e)--(f) Spectral function $\rho_{1\uparrow}(\omega)=\rho_{1\downarrow}(\omega)$ for the nonpolarized model with $t_{\uparrow}=t_{\downarrow}=0.3\Gamma_{N}$ for different side-dot energies: (e) $\epsilon_{2}=0.75\Gamma_{N}$ and (f) $\epsilon_{2}=0$. 
	Other parameters are the same as on panels (a)--(d).
	The inset of panel (f) shows details of the resonant features near the Fermi level for $\epsilon_2=0$.}
}
    \label{fig:corr2}
\end{figure}

In the correlated case ($U_{1} \neq 0$) each Andreev state splits into low and high energy branches separated by energy $U_{1}$, cf., e.g., Ref. \cite{BaranskiJPCM2013}  {for a detailed discussion of this issue}. 
In realistic systems, the Coulomb interactions are usually larger than the energy gap of the superconductor (i.e., $U_{1} \gg \Delta$). 
Consequently, the high energy states coincide with the continuum outside the gap. 
Therefore, high energy branches of Andreev states are beyond considered energy scale and they do not introduce any meaningful physics to the scope of this work.
The detailed analysis of the competition between the local pairing and correlations in the absence of the scattering is conducted in Refs. \cite{Bauer_2007,BaranskiJPCM2013}.   
In Fig.~\ref{fig:corr1}, obtained for strongly correlated regime ($U_{1}=15\Gamma_{N}$), we present the features of the scattering on the background of two low energy Andreev states in the case of spin-polarized tunneling [Fig.~\ref{fig:corr1}(a)] and, for a comparison, in the nonpolarized case [Fig.~\ref{fig:corr1}(b)]. 
One can note that resonant features, described earlier for noncorrelated case, remain qualitatively unchanged despite of strong correlations.
Additionally, one can note a small step near $\omega\approx0$. 
This can be described as underdeveloped Abrikosov-Suhl state also known as Kondo peak. 
These resonances appear as a result of screening of electron spin located on  {the} quantum dot (or impurity) by opposite spin itinerant electrons from the metallic electrode.

To inspect the interplay between scattering features and Kondo state we will analyze the conditions more suitable for full development of the Kondo state. 
Namely, we put energy of QD$_1$ dot slightly below the Fermi level ($\epsilon_{1}=-1.5\Gamma_{N}$), comparable hybridizations ($\Gamma_{S}=\Gamma_{N}$) and low temperature $k_{B}T=0.01\Gamma_{N}$. 
In such conditions, two Andreev states overlap on each other forming a structure resembling Lorentzian. 
One should remember that this structure is still built of two quasiparticle states, which become well-separated if the hybridization to the SC electrode ($\Gamma_S$) is considerably larger than broadening $\Gamma_{N}$ (cf. Fig. 7 of Ref. \cite{BaranskiJPCM2013}).
On the top of that we note well-developed zero-energy Kondo state [cf. Fig.~\ref{fig:corr2}(a)] and one scattering feature for each spin (regular Fano one near $\omega\approx\epsilon_{2}$ for $\rho_{1 \uparrow}(\omega)$ and anomalous Fano one near $\omega\approx-\epsilon_{2}$ in $\rho_{1 \downarrow}(\omega)$].
Panels (a)--(d) of Fig.~\ref{fig:corr2} show what happens when the energy of the side dot gradually approaches to zero, i.e., when the ordinary Fano shape in the spin-$\uparrow$ spectra and the anomalous Fano feature in the spin-$\downarrow$ component overlap with the Kondo state  {at $\omega=0$} [Figs. \ref{fig:corr2}(a), \ref{fig:corr2}(b), \ref{fig:corr2}(c), and \ref{fig:corr2}(d) are obtained for different values of $\epsilon_2$ decreasing from $2.5\Gamma_N$ to $0$]. 
Due to destructive nature of the Fano-like interference the Kondo state in the spin-$\uparrow$ spectra is strongly suppressed when the scattering coincides with the resonant Kondo feature  {[solid blue line in Fig. \ref{fig:corr2}(d)]}. 
In contrast, the anomalous feature  {for} opposite spin electrons seems to enhance the Kondo state  {[dashed red line in Fig. \ref{fig:corr2}(d)]}.

In the realistic model with equal interdot hoppings for electrons of each spin ($t_{\uparrow}=t_{\downarrow}$), both types of electrons are directly scattered (giving the ordinary Fano feature near $\omega\approx\epsilon_2$) as well as coupled with scattered electrons of the opposite spin (giving the anomalous Fano feature near $\omega\approx-\epsilon_2$), cf. Fig. \ref{fig:corr2}(e) as well as Fig.~\ref{fig:corr1}(b). 
Consequently, the ordinary Fano feature originating from direct scattering and the anomalous Fano feature coincide if the energy of the side dot $\epsilon_2$ is equal to $0$. 
As it can be seen in Fig.~\ref{fig:corr2}(f), in such a case, both these resonances also coincide with the Kondo spike. 
We note that, for such parameters, the destructive interference plays a dominant role as zero energy state is strongly suppressed (forming a structure resembling the Fano-Kondo feature). 
Contribution of the anomalous Fano resonance in this case is visible as a small spike slightly below the Fermi level and slight enhancement of the Fano-Kondo feature just above the Fermi level [cf. maximal values in the inset of Fig.~\ref{fig:corr2}(f) and in Fig.~\ref{fig:corr2}(d) for the fully spin-polarized case].
One should note that a similar Fano-Kondo structure was predicted, e.g., for the double-quantum-dot system coupled to ferromagnetic electrodes [c.f., Fig. 4(f) in Ref. \cite{Tao2009}]. However, in that work, the enhancement [and the features presented in the inset of Fig. ~\ref{fig:corr2}(f)] associated with the anomalous Fano resonance are not present there due the absence of the local pairing in the system considered in Ref. \cite{Tao2009}.

\section*{Resonant features in differential conductivity}

\begin{figure}
\includegraphics[width=\size]{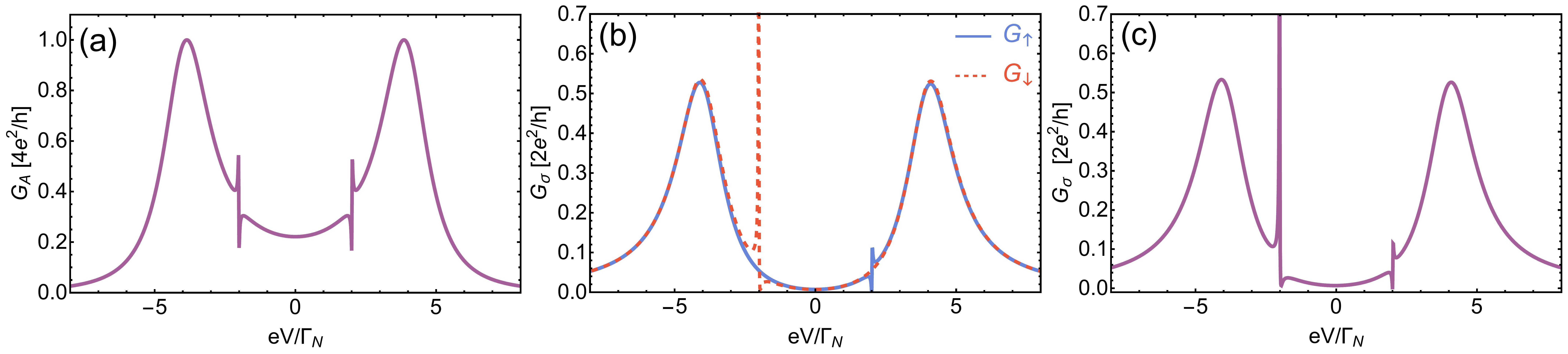}
\caption{
(a) Andreev conductance $G_A=dI_{A}/dV$ and 
(b) single particle conductance $G_{\uparrow}=dI_{\uparrow}/dV$ (solid blue line) and $G_{\downarrow}=dI_{\downarrow}/dV$ (dashed red line), both panels obtained for the spin-polarized case ($t_{\uparrow}=0.3\Gamma_{N}$, $t_{\downarrow}=0$). 
(c) Single particle conductance $G_{\downarrow} = G_ {\uparrow}$ for equal interdot hoppings $t_{\uparrow}=t_{\downarrow}=0.3 \Gamma_{N}$. 
Model parameters used in all three panels are: 
$\Gamma_{S}=4\Gamma_{N}$, $\epsilon_{2}=2 \Gamma_{N}$, $\epsilon_1=0$, $U_{1}=U_{2}=0$, and $k_{B}T=0.01\Gamma_{N}$.}
\label{fig:conduct}
\end{figure}

The spectral function is not  {a} directly measurable quantity. 
Therefore, the resonant features described in this paper can be investigated experimentally only by inspection of differential conductivity $G(V)=dI/dV$. 
For junctions with one metallic and one superconducting electrode low energy charge transport is supported solely by so-called Andreev reflections. 
In such processes, single electron of  {a} given spin from the metallic lead is converted into a Cooper pair propagating in  {the} superconductor with simultaneous reflection of  a hole (with  {the} opposite spin) back to  {the} metal. 
This process, however, involves  {electrons of both spins equally}. 
If an electron of a given spin and of energy $\omega$ is supposed to be converted into a Cooper pair propagating in  {the} superconductor, it needs to ``pick'' additional electron of  {the} opposite spin and of energy $-\omega$. 
Particularly, for energies close to $-\epsilon_{2}$ (where the ``anomalous'' Fano resonance emerges in spectral function of $\sigma=\downarrow$ electrons) electrons are paired with  {the} opposite spin electrons of energy $\epsilon_2$, for which the ordinary Fano resonance emerges.
Consequently, in  {the picture of the Andreev conductivity $G_A=dI_A/dV$}, even for perfect spin-polarized case, resonant features near $\omega\pm \epsilon_{2}$ become a mixture of  {the} ordinary  {Fano} and  {the} ``anomalous'' Fano resonances as seen in Fig. \ref{fig:conduct}(a).  
Indeed, the total Andreev current can be expressed by $I_{A}(V)=\Sigma_{j}I_{A,j}(V)$, where
\begin{eqnarray}
I_{A,j}(V)=\frac{2e^2}{h} \int T_{A,j}(\omega) [ f(\omega - e V)-f(\omega + e V) ] d\omega,
\end{eqnarray}
whereas the Andreev transmittances are given by 
$T_{A,1}(\omega) = \Gamma_{N}^2 \check{G}^{12}_{1 \uparrow}(\omega) $ and 
$T_{A,2}(\omega) = \Gamma_{N}^2 \check{G}^{12}_{1 \downarrow}(\omega)$  
[$\check{G}^{12}_{1 \uparrow}(\omega)$ and $\check{G}^{12}_{1 \downarrow}(\omega)$ are elements of the matrix of Green's functions defined in (\ref{Gr44}); $e$ is the electric charge of  {an} electron and $V$ is the voltage].

To detect  {the} resonant features separately one should rather inspect single particle transport. 
Therefore, we assume that interfacial quantum dot QD$_1$ is connected to an additional metallic electrode. 
Assuming that chemical potential of  {the} superconducting electrode is tuned such, that no current is contributed on average from it (so-called floating lead), the only charge transport left is a single-particle current between two metallic electrodes. 
We calculate single-particle differential conductivity assuming that both normal electrodes are coupled to QD$_1$. 
We assume that  {the} energy unit is equal to a sum of  {the} hybridizations of both metallic electrodes, i.e.,  $\Gamma_{N1}+\Gamma_{N2}=\Gamma_{N}$. 
Single-particle current $I_{\sigma}(V)$ is calculated using the Landauer formula \cite{Landauer57} 
\begin{eqnarray}
I_{\sigma}(V)=\frac{2e^2}{h} \int T_{\sigma}(\omega) [ f(\omega-eV)-f(\omega) ] d\omega,
\end{eqnarray}
where $T_{\sigma}(\omega)=\Gamma_{N}^2 |\check{G}^{11}_{1 \sigma}(\omega)|^{2}$ is a single particle transmittance. 
In Fig. \ref{fig:conduct}, we also present differential conductivity $G_{\sigma}=dI_{\sigma}/dV$ as a function of  {the} applied voltage considering two cases 
(i) the toy model with  {the} perfect spin-polarized scattering [$t_{\uparrow}=0.3\Gamma_{N}$, $t_{\downarrow}=0$, Fig.~\ref{fig:conduct}(b)] and 
(ii) the more realistic case where both spin components can be tunneled between  {the} dots [Fig.~\ref{fig:conduct}(c)]. 
In the first case, the ordinary Fano resonance emerges in  {the} conductivity of directly scattered electrons $\sigma=\uparrow$  [blue line in Fig.~\ref{fig:conduct}(b)], 
while the feature related to pairing with scattered electron is visible as sharp spike in conductance of  {the} opposite spin electron near $\omega\approx-\epsilon_{2}$. 
For the nonpolarized case,  {electrons of each spin $\sigma=\uparrow,\downarrow$} are both directly scattered and bound into a pair with  {the} scattered electron  {of spin $\bar{\sigma}$} [Fig.~\ref{fig:conduct}(c)]. 
Therefore, in  {the} picture of conductivity, we can detect  {the} regular Fano shape near $\omega\approx\epsilon_{2}$ and  {the} anomalous resonant feature near $\omega\approx-\epsilon_{2}$.

\section*{Conclusions}

In the present work, we inspect  the energy spectrum of  the double-quantum-dot system coupled to  a superconducting reservoir in  the T-shape geometry. 
In  the analyzed system, combined effect of  {the} electron scattering and  the local pairing  gives rise to two resonant features on  {the} opposite sides of  {the} Fermi level. 
Considering  the perfectly spin-polarized interdot tunneling regime, we show that one of the resonances  emerges as a result of  the direct scattering. 
The other one  {emerges} as a result of pairing of a given electron with  a scattered electron of  the opposite spin. 
Therefore, an existence of a pair of characteristic Fano and anomalous Fano resonances can be considered as a fingerprints for an occurrence of a bound state in  {the} given system. 
We also obtained characteristics for differential conductivity and identified the features associated with the discussed resonances. 
These results of the work, derived for both strongly asymmetric ($t_{\downarrow}=0$) and symmetric ($t_{\downarrow}=t_{\uparrow}\neq0$) cases, suggest that the regular Fano (near $\omega\approx \epsilon_{2}$) and the anomalous Fano (near $\omega\approx -  \epsilon_{2}$) features could be detected in real nanoscopic systems.
Such resonances can be observed in  {a} variety of complex nano-systems (coupled to a superconductor) where  {the} broadening of energy levels for each subsystem is considerably different. 
Although the spin-polarized model is hardly achievable experimentally without using the magnetic field, it allows to uncover the mechanism behind the formation of resonant features on both sides of the Fermi level appearing also in the realistic nonpolarized model (i.e., the symmetric case of $t_{\uparrow}=t_{\downarrow}$) \cite{BaranskiPRB2011,BaranskiPRB2012,Glodzik2017}.
Note also that systems in such a configuration (i.e., T-shape one) can be investigated experimentally (cf. Refs. \cite{FERNANDES201898,TorioPRB2002,Orellana2003,Takazawa_DQDT,KangDQD,%
BulkaPRL2001,GUEVARA2006,WeymannPRB2014,GramichPRB2017,Bordoloi2019} and references therein).

In this work, we showed that the resonant feature that  originates from direct scattering can be described in terms of the Fano-like function with great details. 
Particularly, for  the double-quantum-dot system coupled to a (normal) metal and a superconductor (Fig. \ref{fig:1.system}), the spectral function of directly scattered electrons was approximated by product of the Fano line-shape and the Andreev states. 
A convergence of such approximation and the exact spectral function turned out to be very accurate. 
To achieve satisfactory convergence for  {the} resonant feature on  {the} opposite side of the Fermi level, one needed to impose additional correction $(\phi_0)$ to the Fano function  {[Eq. (\ref{eq:rstardown})]}. 
Using such an assumption, we managed to achieve  {a} good convergence,  {but ``correction''} parameter $\phi_0$ becomes enormously large, when  {the} resonant level approaches  {the} Fermi surface [e.g., for $|\epsilon_{2}|<0.1\Gamma_N$, parameter $\phi_0$ becomes two order of magnitude higher than assumed energy unit (i.e., $\phi_{0} \approx 300\Gamma_{N}$), also the asymmetry parameter in such cases becomes as large as $q_{\downarrow} \approx 80$]. 
Therefore, we proposed to approximate such resonances by the inversion of the Fano function  {[Eq. (\ref{eq:rtilde})]} rather than the direct Fano one. 
Using this assumption, we achieved a high convergence with keeping correction $\tilde{\phi}_{0}$ small for a wide range of  {the} model parameters. 
We also discussed  the interplay of both Fano-like features with the Kondo resonance in the presence of correlations.

\bibliography{biblio_QDOTS4}

\section*{Acknowledgements}

The authors express their sincere thanks to Tadeusz Doma\'nski and Andrzej Ptok for useful discussions on some issues raised during preparation of this work.
The research has been conducted in the framework of the project implemented in 2018-2021, entitled  ``Analysis of nanoscopic systems coupled with superconductors in the context of  quantum information processing''  no. GB/5/2018/209/2018/DA  funded by the Ministry of National Defence, Republic of Poland (J.B. and T.Z.), as well as National Science Centre (NCN), Poland, under grant SONATINA no. UMO-2017/24/C/ST3/00276 (M.B. and K.J.K.) in years 2017-2020.

\section*{Author contributions statement}

J.B. and K.J.K. contributed equally to posing the problem.
J.B. initialized and coordinated the project. 
J.B., T.Z., and M.B. derived the analytic expressions.  
J.B. and K.J.K. performed numerical calculations. 
All authors consulted the obtained results and contributed to the discussions and analysis of the results.
J.B. and K.J.K. prepared the first version of the manuscript. 
All authors reviewed the manuscript.
J.B. and K.J.K. wrote the paper in its final form. 
All authors accepted it.

\section*{Additional information}

\textbf{Competing Interests:} 
The authors declare no competing interests. 
The mentioned funding sources had no role in the design of this study and did not have any role during its execution, analyses, interpretation of the data, or decision to submit results.

\section*{}

\section*{Information about the published article}

\textbf{This is author-created e-print version of the article.}
This manuscript was submitted to Scientific Reports journal (publisher: Springer Nature / Nature Research) and it was published there (Received: 14 October 2019; Accepted: 22 January 2020; Published: 19 February 2020) under a Creative Commons Attribution 4.0 International License, which permits use, sharing, adaptation, distribution and reproduction in any medium or format, as long as you give appropriate credit to the original author(s) and the source, provide a link to the Creative Commons license, and indicate if changes were made. 
To view a copy of this license, visit \url{http://creativecommons.org/licenses/by/4.0/}.\\
$\quad$ \\ 
\textbf{The original article should be cited as}: \\
$\quad$ \\
Bara\'nski, J., Zienkiewicz, T., Bara\'nska, M., and Kapcia, K. J.: \\
\emph{Anomalous Fano Resonance in Double Quantum Dot System Coupled to Superconductor}. \\
Scientific Reports \textbf{10}, 2881 (2020). \\
DOI: 10.1038/s41598-020-59498-y; URL: \url{https://doi.org/10.1038/s41598-020-59498-y}

\end{document}